\documentstyle[aps,12pt]{revtex}

\begin{document}
\author{E.N. Glass\thanks{%
Permanent address: Physics Department, University of Windsor, Ontario N9B
3P4, Canada} and J.P. Krisch}
\address{Department of Physics, University of Michigan, Ann Arbor, Michgan 48109}
\date{18 May 2000}
\title{Fractal Scales in a Schwarzschild Atmosphere}
\maketitle

\begin{abstract}
\newline
\newline
Recently, Glass and Krisch have extended the Vaidya radiating metric to
include both a radiation fluid and a string fluid [1999 Class. Quantum Grav. 
{\bf 16}, 1175]. Mass diffusion in the extended Schwarzschild atmosphere was
studied. The continuous solutions of classical diffusive transport are
believed to describe the envelope of underlying fractal behavior. In this
work we examine the classical picture at scales on which fractal behavior
might be evident.\newline
\newpage\ 
\end{abstract}

\section{INTRODUCTION}

Since its discovery in 1916, the Schwarzschild vacuum solution has been a
source of many insights into relativistic physics. 
\begin{equation}
g_{ab}^{Sch}dx^adx^b=(1-2m_0/r)du^2+2dudr-r^2d\Omega ^2.  \label{sch-met}
\end{equation}

This solution and its extensions contain new views of relativistic physics
on a variety of scales, from the size of the Universe down to the size of a
stellar black hole. When contemplating quantum effects at classical
boundaries such as the Schwarzschild horizon, the scales become much smaller
down to the Planck length $L_p=\sqrt{\hbar G/c^3}=1.6\times 10^{-33}$ $cm$.
The classical point particle electron may perhaps be resolved into Planck
scale strings bits with size proportional to the Planck length and with $%
10^{20}$ strings bits fitting into a classical electron radius.

Vaidya \cite{vaidya} extended the solution by allowing the Schwarzschild
mass parameter to be a function of retarded time, $m(u)$, and discovered a
spherically symmetric null fluid atmosphere beyond the horizon at $2m_{0}$.
Glass and Krisch (GK) \cite{ed-jean2} pointed out that allowing the
Schwarzschild mass parameter to be a function of radial position and
retarded time, $m(u,r)$, creates an atmosphere with both a null fluid and
string fluid stress-energy content described by metric $g_{ab}^{GK}$. The
density of the string fluid, $\rho $, obeys a classical diffusion equation
in the Schwarzschild spacetime: 
\begin{equation}
\partial _{u}\rho =r^{-2}\partial _{r}(Dr^{2}\partial _{r}\rho )
\label{diff1}
\end{equation}
where $D(r)$ is the variable diffusivity \cite{gh-lang} found in anomalous
diffusion. One may show \cite{ed-jean3}, if the mass is diffusing in a
flowspace with $\sqrt{-g}=r^{-\beta }f(\vartheta )$ (in \cite{ed-jean3} $%
\beta =1-\delta $), that the functional dependence of the diffusivity is
determined. 
\begin{eqnarray}
\partial _{u}m &=&r^{\beta }\partial _{r}(Dr^{-\beta }\partial _{r}m)
\label{diff2} \\
D(r) &=&D_{_{0}}r^{\beta -2}.  \nonumber
\end{eqnarray}
The continuous solutions of classical diffusive transport are believed to
describe the envelope of underlying fractal behavior \cite{osh-pro}. Fractal
behavior is usually associated with mass scaling but the radial dependence
of the diffusivity \cite{a-o} carries additional information about the
connectivity of the underlying fractal substrate, and so we wish to examine
the classical picture at scales on which fractal behavior might be evident.
These smaller scales are still classical (i.e. larger than quantum scales).
We consider a 2-parameter family of spacetimes with metric function $%
A=1-2m(u,r)/r$ 
\begin{equation}
g_{ab}^{scale}dx^{a}dx^{b}=Adu^{2}+2dudr-\lambda ^{2-\alpha }r^{\alpha
}d\Omega ^{2},  \label{frac-met}
\end{equation}
where $\lambda $ has units of length, and is required for spheres to have
area of (length)$^{2}$. This metric is capable of describing different
scales for a particular central source. For example, a particular mass
function $m(u,r)$ with $\alpha =2$, could describe the Schwarzschild vacuum
outside the source. For $\alpha \neq 2$ it would describe the source at
different scales where we expect evidence of fractal behavior to appear.
Fractal dimensions are generally less than the Euclidean dimension, and we
will describe an averaging process on 2-spheres for $\alpha \leq 2$. The
scale parameter $\alpha $ cannot be removed by a coordinate transformation.
If one attempts the appearance of the usual area coordinate by $%
y^{2}=\lambda ^{2-\alpha }r^{\alpha }$, then $\alpha $ will emerge in the ($%
u $,$r$) part of $g_{ab}^{scale}$. Moreover, the manifestly
coordinate-independent sectional curvature of the ($\vartheta ,\varphi $)
two-surfaces, given in Eq.(\ref{mdef}) below, provides a physically
meaningful mass expression in terms of $\alpha $.

The interpretation of $\alpha $ as a fractal scale follows from considering
diffusive processes in the atmosphere. If ${\cal P}(u,r)$ is a quantity that
is diffusing, the diffusion equation is 
\begin{equation}
\partial _{u}{\cal P}=r^{-\alpha }\partial _{r}(Dr^{\alpha }\partial _{r}%
{\cal P})  \label{p-eqn}
\end{equation}
and $\alpha +1$ is identified as the dimension of the fractal substrate on
which the diffusion is occurring \cite{osh-pro},\cite{d-m-s}. At smaller
quantum levels, ${\cal P}$ can be a probability measure on the space of
radial paths \cite{nels}.

The fluid atmosphere described by this metric is discussed in the next
section. In section III diffusion and scale sizes are discussed. Definitions
of topological and Hausdorff dimensions are given. Applications of the
fractal view are presented in section IV, along with specific mass
solutions. The paper is summarized in the final section.

In this work Latin indices range over ($0$,$1$,$2$,$3$) = ($u,r,\vartheta
,\varphi $). Our sign conventions are $2A_{c;[ab]}=A_eR_{\ \ cab}^e,$ and $%
R_{ab}=R_{\ abc}^c.$ $\ $Overdots abbreviate $\partial /\partial u$, and
primes abbreviate $\partial /\partial r$. Overhead carets denote unit
vectors. We use units where $G=c=1$. Einstein's field equations are $%
G_{ab}=-8\pi T_{ab},$ and the metric signature is (+,-,-,-).

\section{THE ATMOSPHERE}

Metric $g_{ab}^{scale}$ describes an atmosphere that admits a two-fluid
description of matter with outward flowing short-wavelength photons
(sometimes called a ''null fluid'').

With the use of a Newman-Penrose null tetrad the Einstein tensor for the
atmosphere is computed from (\ref{frac-met}) and given by 
\begin{eqnarray}
G_{ab} &=&-2\Phi _{11}(l_{a}n_{b}+n_{a}l_{b}+m_{a}\bar{m}_{b}+\bar{m}%
_{a}m_{b})  \label{ein1} \\
&&-2\Phi _{00}n_{a}n_{b}-2\Phi _{22}l_{a}l_{b}-6\Lambda g_{ab}.  \nonumber
\end{eqnarray}
Here the null tetrad components of the Ricci tensor are 
\begin{mathletters}
\begin{eqnarray}
\Phi _{00} &=&\frac{\alpha (2-\alpha )}{4r^{2}},  \label{ein1a} \\
\Phi _{11} &=&\frac{2m^{\prime }-rm^{\prime \prime }}{4r^{2}}-\frac{\alpha
^{2}}{16r^{2}}+\frac{\lambda ^{\alpha -2}}{4r^{\alpha }}+\frac{(\alpha
-2)(\alpha +2)m}{8r^{3}},  \label{ein1b} \\
\Phi _{22} &=&-\frac{\alpha \dot{m}}{2r^{2}}+\frac{\alpha (2-\alpha )}{%
16r^{2}}(1-2m/r)^{2},  \label{ein1c} \\
\Lambda &=&R/24=\frac{rm^{\prime \prime }+2(\alpha -1)m^{\prime }}{12r^{2}}+%
\frac{(\alpha -2)(3\alpha -2)m}{24r^{3}}  \label{ein1d} \\
&&-\frac{\alpha (3\alpha -4)}{48r^{2}}+\frac{\lambda ^{\alpha -2}}{%
12r^{\alpha }}.  \nonumber
\end{eqnarray}
The only non-zero component of the Weyl tensor is 
\end{mathletters}
\begin{equation}
\Psi _{2}=-\frac{(\alpha +1)m}{3r^{3}}+\frac{(2+\alpha )m^{\prime
}-rm^{\prime \prime }}{6r^{2}}+\frac{\alpha }{12r^{2}}-\frac{\lambda
^{\alpha -2}}{6r^{\alpha }}.  \label{psi2}
\end{equation}
The metric is Petrov type {\bf D} with $l^{a}$ and $n^{a}$ principal null
geodesic vectors 
\begin{mathletters}
\begin{eqnarray}
l_{a}dx^{a} &=&du,\ \ \ \ \ \ \ \ \ \ \ \ \ \ \ \ \ \ \ \ \ \ \ \ \ \ \ \ \
\ \ \ \ \ \ \ \ \ \ \ \ \ \ \ \ l^{a}\partial _{a}=\partial _{r}
\label{nteta} \\
n_{a}dx^{a} &=&(A/2)du+dr,\ \ \ \ \ \ \ \ \ \ \ \ \ \ \ \ \ \ \ \ \ \ \ \ \
\ \ \ \ \ n^{a}\partial _{a}=\partial _{u}-(A/2)\partial _{r}  \label{ntetb}
\\
m_{a}dx^{a} &=&-(1/\surd 2)\lambda ^{1-\alpha /2}r^{\alpha /2}(d\vartheta
+i\ \text{sin}\vartheta d\varphi ),\ \ m^{a}\partial _{a}=(1/\sqrt{2}%
)\lambda ^{\alpha /2-1}r^{-\alpha /2}(\partial _{\vartheta }+\frac{i}{\text{%
sin}\vartheta }\partial _{\varphi }).  \label{ntetc}
\end{eqnarray}
When $\alpha =2$ all quantities have the spacetime values for metric $%
g_{ab}^{GK}$ discussed in \cite{ed-jean2}.

In order to clearly see the matter content we introduce a time-like unit
velocity vector $\hat{v}^{a}$ and three unit space-like vectors $\hat{r}^{a}$%
, $\hat{\vartheta}^{a}$, $\hat{\varphi}^{a}$ such that 
\end{mathletters}
\[
g_{ab}^{scale}=\hat{v}_{a}\hat{v}_{b}-\hat{r}_{a}\hat{r}_{b}-\hat{\vartheta}%
_{a}\hat{\vartheta}_{b}-\hat{\varphi}_{a}\hat{\varphi}_{b}. 
\]
The unit vectors are defined by 
\begin{mathletters}
\begin{eqnarray}
\hat{v}_{a}dx^{a} &=&A^{1/2}du+A^{-1/2}dr,\ \ \ \ \hat{v}^{a}\partial
_{a}=A^{-1/2}\partial _{u},  \label{vteta} \\
\hat{r}_{a}dxa &=&A^{-1/2}dr,\ \ \ \ \ \ \ \ \ \ \ \ \ \ \ \hat{r}%
^{a}\partial _{a}=A^{-1/2}\partial _{u}-A^{1/2}\partial _{r},  \label{vtetb}
\\
\hat{\vartheta}_{a}dx^{a} &=&\lambda ^{1-\alpha /2}r^{\alpha /2}d\vartheta
,\ \ \ \ \ \ \ \ \ \ \hat{\vartheta}^{a}\partial _{a}=-\lambda ^{\alpha
/2-1}r^{-\alpha /2}\partial _{\vartheta },  \label{vtetc} \\
\hat{\varphi}_{a}dx^{a} &=&\lambda ^{1-\alpha /2}r^{\alpha /2}\text{sin}%
\vartheta d\varphi ,\ \ \ \ \hat{\varphi}^{a}\partial _{a}=-\lambda ^{\alpha
/2-1}(r^{\alpha /2}\text{sin}\vartheta )^{-1}\partial _{\varphi }.
\label{vtetd}
\end{eqnarray}

The Einstein tensor (\ref{ein1}) yields the energy-momentum tensor 
\end{mathletters}
\begin{equation}
T_{ab}=\psi l_{a}l_{b}+\rho \hat{v}_{a}\hat{v}_{b}+p_{r}\hat{r}_{a}\hat{r}%
_{b}+p_{\perp }(\hat{\vartheta}_{a}\hat{\vartheta}_{b}+\hat{\varphi}_{a}\hat{%
\varphi}_{b})+q_{a}\hat{v}_{b}+\hat{v}_{a}q_{b}  \label{energy-mom1}
\end{equation}
where 
\begin{mathletters}
\begin{eqnarray}
8\pi \psi &=&-\frac{\alpha \dot{m}}{r^{2}}+\frac{\alpha (2-\alpha )A^{2}}{%
8r^{2}},  \label{em-a} \\
8\pi \rho &=&\frac{\alpha m^{\prime }}{r^{2}}+\frac{\lambda ^{\alpha -2}}{%
r^{\alpha }}-\frac{\alpha (\alpha -1)}{2r^{2}}+\frac{(\alpha -2)}{8r^{3}}%
[8\alpha m-\alpha (r-2m)],  \label{em-b} \\
8\pi p_{r} &=&-\frac{\alpha m^{\prime }}{r^{2}}-\frac{\lambda ^{\alpha -2}}{%
r^{\alpha }}+\frac{\alpha (\alpha -1)}{2r^{2}}-\frac{(\alpha -2)}{8r^{3}}%
[8\alpha m+\alpha (r-2m)],  \label{em-c} \\
8\pi p_{\perp } &=&-\frac{m^{\prime \prime }}{r}-\frac{(\alpha -2)m^{\prime }%
}{r^{2}}-\frac{(\alpha -2)^{2}m}{2r^{3}}+\frac{\alpha (\alpha -2)}{4r^{2}},
\label{em-d} \\
8\pi q_{a} &=&[A\frac{\alpha (2-\alpha )}{8r^{2}}]\,\hat{r}_{a}.
\label{em-e}
\end{eqnarray}
The string equation of state $\rho +p_{r}=0$ holds for the scale with $%
\alpha =2$. The decomposition of the energy-momentum tensor given in Eq.(\ref
{energy-mom1}) is less compelling on $\alpha \neq 2$ scales since there is
heat flow in addition to the null fluid. The heat flow arises from the $\Phi
_{00}n_{a}n_{b}$ term in $G_{ab}$. This is seen only on $\alpha \neq 2$
scales and is the intermediate electromagnetic field part of the null fluid,
which has only a far field part in the usual Vaidya large scale view. A
different covariant decomposition using $l_{a}dx^{a}=A^{-1}(\hat{v}_{a}-\hat{%
r}_{a})dx^{a}$ provides a single fluid with radial heat flow. The
energy-momentum tensor decomposes as 
\end{mathletters}
\begin{equation}
T_{ab}=\tilde{\rho}\hat{v}_{a}\hat{v}_{b}+\tilde{p}_{r}\hat{r}_{a}\hat{r}%
_{b}+p_{\perp }(\hat{\vartheta}_{a}\hat{\vartheta}_{b}+\hat{\varphi}_{a}\hat{%
\varphi}_{b})+\tilde{q}_{a}\hat{v}_{b}+\hat{v}_{a}\tilde{q}_{b}
\label{energy-mom2}
\end{equation}
where 
\begin{mathletters}
\label{em2-comps}
\begin{eqnarray}
8\pi \tilde{\rho} &=&-A^{-1}\frac{\alpha \dot{m}}{r^{2}}+\frac{\alpha
m^{\prime }}{r^{2}}+\frac{\lambda ^{\alpha -2}}{r^{\alpha }}-\frac{\alpha
(\alpha -1)}{2r^{2}}+\frac{(\alpha -2)}{4r^{3}}[4\alpha m-\alpha (r-2m)], \\
8\pi \tilde{p}_{r} &=&-A^{-1}\frac{\alpha \dot{m}}{r^{2}}-\frac{\alpha
m^{\prime }}{r^{2}}-\frac{\lambda ^{\alpha -2}}{r^{\alpha }}+\frac{\alpha
(\alpha -1)}{2r^{2}}-\frac{(\alpha -2)}{4r^{3}}[4\alpha m+\alpha (r-2m)], \\
8\pi \tilde{q}_{a} &=&(A^{-1}\frac{\alpha \dot{m}}{r^{2}})\,\hat{r}_{a}.
\end{eqnarray}
We note that 
\end{mathletters}
\[
8\pi (\tilde{\rho}+\tilde{p}_{r})=-A^{-1}\frac{2\alpha \dot{m}}{r^{2}}-\frac{%
\alpha (\alpha -2)}{2r^{3}}(r-2m). 
\]
In the static case the string equation of state $\tilde{\rho}+\tilde{p}%
_{r}=0 $ holds on the horizon for all $\alpha $. For mass $m(r)$ which falls
off at least as fast as $O(1/r^{2})$, the string equation of state also
holds at future null infinity.

The metric function $m(u,r)$ measures the system mass only for $\alpha =2$,
the GK metric. In general, spherical symmetry allows the mass within
two-surfaces of constant $u$ and $r$ to be invariantly defined by the
sectional curvature \cite{m-s} of those surfaces: 
\begin{eqnarray}
-2M/r^{3} &:&=R_{abcd}\hat{\vartheta}^{a}\hat{\varphi}^{b}\hat{\vartheta}^{c}%
\hat{\varphi}^{d}  \label{mdef} \\
&=&-\frac{\alpha ^{2}m}{2r^{3}}+\frac{\alpha ^{2}}{4r^{2}}-\frac{\lambda
^{\alpha -2}}{r^{\alpha }}.  \nonumber
\end{eqnarray}
When $\alpha =2$ then $M=m$.

\section{DIFFUSION AND SCALE}

\subsection{Diffusion}

Continuous diffusive processes can envelope underlying fractal behavior. In
the $\alpha =2$ Schwarzschild atmosphere, the density $\rho $ and the mass
function $m$ diffuse according to Eqs.(\ref{diff1}) and (\ref{diff2}). By
analogy with classical diffusion of density in the Schwarzschild spacetime,
we assume that the density enveloping fractal behavior obeys 
\begin{equation}
\partial _{u}\Gamma =r^{-\alpha }\partial _{r}(Dr^{\alpha }\partial
_{r}\Gamma ),  \label{gam-diff}
\end{equation}
where the mass density $\Gamma $ is defined on $u=const$ surfaces, using the
sectional curvature mass, by the relation $M=\int \Gamma \sqrt{-g}d^{3}x$ so
that 
\begin{equation}
4\pi (\Gamma -\Gamma _{0}):=\lambda ^{\alpha -2}r^{-\alpha }\partial _{r}M.
\label{gam-def}
\end{equation}
The relationship between the energy density $\rho ,$ seen by observer $\hat{v%
}^{a},$ in (\ref{em-b}) and the mass density is then given by 
\begin{eqnarray}
4\pi \Gamma &=&4\pi \rho +(\frac{1}{8r^{2}})[(\lambda /r)^{\alpha -2}\alpha
^{2}(2m^{\prime }-1)-2\alpha (2m^{\prime }-\alpha +1)]  \label{rho-gam} \\
&&+(\frac{1}{2r^{2}})(\lambda /r)^{\alpha -2}[(3-\alpha )(\lambda
/r)^{\alpha -2}-1]-\frac{(\alpha -2)}{16r^{3}}[8\alpha m-\alpha (r-2m)]. 
\nonumber
\end{eqnarray}
In the region where $\alpha =2$, the two densities $\Gamma $ and $\rho $ are
equal.

Assumption (\ref{gam-diff}) is equivalent to assuming a Fickian mass current
of the form 
\begin{equation}
\partial _{u}M=4\pi D(r)\,r^{\alpha }\partial _{r}\Gamma ,  \label{fick-cur}
\end{equation}
and then $\partial _{r}M=4\pi \lambda ^{2-\alpha }r^{\alpha }(\Gamma -\Gamma
_{0})$ yields (\ref{gam-diff}). Furthermore, if the sectional curvature mass 
$M$ diffuses as 
\begin{equation}
\partial _{u}M=r^{\beta }\partial _{r}(Dr^{-\beta }\partial _{r}M)
\label{m2-diff}
\end{equation}
then $D(r)$ is determined by $\partial _{u}M$ in (\ref{fick-cur}) and $%
\partial _{r}M$: 
\begin{equation}
D(r)=D_{_{0}}r^{\beta -\alpha }.  \label{d-func}
\end{equation}
Note\ that\ if\ $\alpha =2$ the $M$ diffusion equation is just Eq.(\ref
{diff2}) with $\beta =1-\delta .$ For\ any\ $\alpha $, when $D=D_{_{0}}$, it
follows that\ $\beta =\alpha $ and that $M$\ and\ $\Gamma $ diffuse\
reciprocally.

Equation (\ref{m2-diff}) admits the homogeneous solution 
\begin{equation}
M_{\hom }(r)=M_{0}+M_{1}(r/\lambda )^{1+\alpha }  \label{m-homo}
\end{equation}
which can be added to each time-dependent solution. There are two important
observations that follow from these results. First, the metric of $M_{\hom }$
is not asymptotically flat and the diffusive solution would need to be
matched to a solution with appropriate asymptotic behavior. Second, the
diffusivity depends on a coordinate and not on the size of any correlated
mass grouping. This implies that there may be multiple distance scales in
the atmosphere.

\subsection{Dimensions}

To better understand fractal scales we briefly discuss fractal dimensions 
\cite{hughes}. We cannot use the intuitive notion of the number of
coordinates needed to locate a point in some region as the dimension of the
region, since Peano constructed a curve ($1$ coordinate) which fills a
square ($2$ coordinates). Consider a point in set $\Omega $, with $\Omega $
a subset of an Euclidean space of integer dimension $d_{E}$. The notion of
topological dimension involves the idea that neighborhoods around a point
have boundaries of smaller dimension. Precisely, the topological dimension $%
d_{T}(\Omega )$ of set $\Omega $ is the smallest integer such that each
point has small neighborhoods with boundaries of lesser dimension. Since $%
d_{T}(\emptyset )=-1$ it is clear that $d_{T}(\Omega )\leq d_{E}$.

The Hausdorff dimension \cite{haus} of self-similar sets (fractals) is given
in terms of coverings by open sets and is defined as 
\[
d_H(\Omega ):=\lim_{r\rightarrow 0}\frac{-\ln N(r)}{\ln r} 
\]
where $N(r)$ is the smallest number of open balls needed to cover $\Omega $.
The definition of $d_H$ is equivalent to the power law $N(r)\approx
const\times r^{-d_H}.$ When $d_H=3$, $N(r)$ corresponds to a 3-dimensional
Newtonian specific density.

The Hausdorff dimension of a fractal set is not always an integer. The
scaling dimension $d_{S}$ of fractal sets is more intuitive and, since
geometric fractals display regular self-similarity, is equal to the
Hausdorff dimension: 
\[
d_{S}(\Omega )=\frac{\log \text{(number of replicas)\ }}{\text{%
log(magnification factor)}}. 
\]
Two famous examples are $d_{S}($Cantor set$)=\frac{\log (2)}{\log (3)}$ in $%
d_{E}=1$, and $d_{S}($Sierpinski gasket$)=\frac{\text{log}(3)}{\text{log}(2)}
$ in $d_{E}=2$. An integer valued example is $d_{S}($Sierpinski gasket$)=%
\frac{\log (4)}{\log (2)}$ in $d_{E}=3,$ since all Sierpinski gaskets obey $%
d_{S}(\Omega )=\frac{\log \text{(}d_{E}+1\text{)}}{\text{log(}2\text{)}}$. \
In general, $d_{T}(\Omega )\leq d_{H}(\Omega )\leq d_{E}$.

\subsection{Scales}

Maximum scale size is independent of any particular application. Since $%
\alpha +1$ is the dimension of the fractal substrate, bounded by $d_{E}$, $%
\alpha _{\max }=2$.

Consider a vacuum solution at $\alpha =2$ and ask how it is produced.
Beginning at a small scale, one considers the atmosphere to be made of many
regions. Within each region an average energy density is found; in some of
the regions it will be negative, in some positive, and in some zero. Then a
larger region, containing several small regions is considered and again an
average energy density is produced until in the end we have the vacuum
atmosphere with all of the fluctuations averaging to zero. This process is
just a renormalization group averaging procedure and, just as in
renormalization group calculations on a lattice, it is necessary at each
averaging step to rescale. In lattice calculations, one rescales to maintain
the lattice spacing. In an atmospheric calculation with a horizon, the
rescaling is done to always keep the physical observers outside the horizon.
There are several ways in which this could be accomplished.

One method is to have an observer at constant $r$ examine an area of his
2-sphere at different scales while maintaining the same coordinate value of $%
r$ at all scales. The entire 2-sphere at $\alpha =2$ can be tiled by
elements of area $\delta A_{2}.$ The sum of all the tiles is the total area
of the sphere $4\pi r^{2}$. At small scales, the observer chooses a region
of the sphere containing many area elements to average over. The result of
this averaging must be rescaled at each step, as in lattice renormalization,
so that the rescaled area is again $\delta A_{2}$. With this averaging
procedure, an observer outside the horizon for $\alpha =2$ will remain
outside of the horizon for all $\alpha $. The $\alpha =2$ horizon position
will persist at all scales.

To see the effect of the rescaling, consider a typical averaging process.
Assume that $\alpha =2$ is the vacuum solution and let the observer look at
an area $\delta A_{2}$ and solid angle $\delta \Omega _{2}$ with $\delta
A_{2}=r^{2}\delta \Omega _{2}.$ This region is the result of averaging over
larger regions at the lower scale and then rescaling the area size. The
larger region would have an area $\delta A_{\alpha }=\lambda ^{2-\alpha
}r^{\alpha }\delta \Omega _{\alpha }.$ The rescaling would adjust $\delta
\Omega _{\alpha }$ to the size of $\delta \Omega _{2\text{ }}$and thereby
rescale the areas. The rescaling formulas are 
\begin{eqnarray}
(\lambda /r)^{2-\alpha }\delta \Omega _{\alpha } &\longmapsto &\delta \Omega
_{2}  \label{rescale} \\
\delta A_{\alpha } &\longmapsto &\delta A_{2}.  \nonumber
\end{eqnarray}
The scaling factor $(\lambda /r)^{2-\alpha }$ is expected to be $\leq 1$.
For metrics without a horizon, the scale factor can be chosen. When there is
a horizon, the scale factor is determined by its value on the horizon. To
explore the numerical values of this factor in metrics with a horizon, a
specific mass solution is needed.

\section{APPLICATIONS}

\subsection{Constant mass solution}

A specific mass solution is necessary in order to explicitly treat diffusion
and fractal behavior. We have seen above that the sectional curvature mass
can be written as 
\begin{equation}
2M=\frac{\alpha ^{2}}{4}(2m-r)+\lambda ^{\alpha -2}r^{3-\alpha }.
\label{mdef2}
\end{equation}
To illustrate the rescaling process, we explore the case of constant $M$.
For $\alpha =2$ there is a Schwarzschild vacuum with an horizon at $\
r=2m_{0}$. Assuming that all scales see the horizon at $2m_{0}$ we can
determine the scaling factor. From Eq.(\ref{mdef2}) with $M=m_{0}$, and $%
r=2m=2m_{0}$ it follows that 
\[
(2m_{0}/\lambda )^{\alpha -2}=1 
\]
or $\lambda =2m_{0}$.

With $m(r)$ from Eq.(\ref{mdef2}) the fluid parameters in Eq.(\ref{em2-comps}%
) are 
\begin{mathletters}
\label{mchoice}
\begin{eqnarray}
8\pi \tilde{q}_a &=&0, \\
8\pi \tilde{\rho} &=&\frac{6(\alpha -2)}\alpha (m_0r^{-3}), \\
8\pi \tilde{p}_r &=&-\frac{2(\alpha -2)}\alpha [m_0r^{-3}+\lambda
^{-2}(\lambda /r)^\alpha ], \\
8\pi p_{\perp } &=&(\frac{\alpha -2}\alpha )^2[-2m_0r^{-3}+\lambda
^{-2}(\lambda /r)^\alpha ].
\end{eqnarray}
For this simple example, $\tilde{\rho}$, resulting from fluctuations, is
negative \cite{f-r} and there is no bounding $\tilde{p}_r=0$ surface. The
radial profile describes an atmosphere asymptotically approaching vacuum. If
one examines the $\alpha <2$ metric, it appears not to be asymptotically
flat. However, asymptotic flatness applies only to the observer's $\alpha =2$
spacetime.

\subsection{Solution with finite boundaries}

The $M=m_{0}$ solution is part of the homogeneous solution, Eq.(\ref{m-homo}%
). The complete homogeneous solution, modified to produce vacuum
Schwarzschild with a horizon at $r=2m_{0}$ is 
\end{mathletters}
\begin{equation}
M=m_{0}+m_{1}(2-\alpha )(r/\lambda )^{\alpha +1}  \label{m-homo2}
\end{equation}
with 
\begin{equation}
m=\frac{4m_{0}}{\alpha ^{2}}+\frac{4m_{1}(2-\alpha )}{\alpha ^{2}}(r/\lambda
)^{\alpha +1}+\frac{r}{2}-\frac{2\lambda }{\alpha ^{2}}(r/\lambda
)^{3-\alpha }.  \label{hor-m2}
\end{equation}
Evaluating this equation on the horizon at $r=2m_{0}$ to find the scaling
factors gives the equation, with $x=\lambda /2m_{0}$, 
\begin{equation}
1+(m_{1}/m_{0})(2-\alpha )x^{-(\alpha +1)}-x^{\alpha -2}=0.  \label{x-eqn}
\end{equation}
This more complex scaling equation has no solution for $m_{1}\geq m_{0}$,
and may describe solutions without an horizon. For $m_{1}<<m_{0}$ there are
two scaling factors. Some typical values for $m_{1}/m_{0}=0.01$ are 
\begin{eqnarray}
\alpha &=&7/4,\ \ \ x\simeq 1,\ 1/7.6  \label{roots-x-eqn} \\
\alpha &=&6/4,\ \ \ x\simeq 1,\ 1/11.9  \nonumber \\
\alpha &=&5/4,\ \ \ x\simeq 1,\ 1/24.5  \nonumber \\
\alpha &=&1,\text{ \ }\ \ \ \text{ }x\simeq 1,\ 1/99  \nonumber
\end{eqnarray}
For the complete homogeneous solution the fluid parameters are 
\begin{mathletters}
\label{mchoice2}
\begin{eqnarray}
8\pi \tilde{q}_{a} &=&0, \\
8\pi \tilde{\rho} &=&\frac{2(\alpha -2)}{\alpha }[3m_{0}r^{-3}+(4-5\alpha
)(m_{1}/\lambda ^{3})(r/\lambda )^{\alpha -2}], \\
8\pi \tilde{p}_{r} &=&-\frac{2(\alpha -2)}{\alpha }[m_{0}r^{-3}+\lambda
^{-2}(\lambda /r)^{\alpha }-3\alpha (m_{1}/\lambda ^{3})(r/\lambda )^{\alpha
-2}], \\
8\pi p_{\perp } &=&(\frac{\alpha -2}{\alpha })^{2}[-2m_{0}r^{-3}+\lambda
^{-2}(\lambda /r)^{\alpha }]+(\frac{\alpha -2}{\alpha })[(5\alpha
-4)(2m_{1}/r^{3})(r/\lambda )^{\alpha +1}].
\end{eqnarray}
For solution (\ref{hor-m2}) there are boundaries for some $\alpha .$ The
boundary radii follow from $\tilde{p}_{r}=0$ at $\lambda =2m_{0}$. With $%
y=r_{b}/2m_{0}$, 
\end{mathletters}
\begin{equation}
1+2y^{3-\alpha }-3\alpha (m_{1}/m_{0})y^{\alpha +1}=0.  \label{y-eqn}
\end{equation}
Again choosing $m_{1}/m_{0}=0.01,$ there are boundaries which get larger as $%
\alpha $ decreases: 
\begin{eqnarray}
\alpha &=&7/4,\ \ \ y\simeq 11.5\   \label{roots-y-eqn} \\
\alpha &=&6/4,\ \ \ y\simeq 45\   \nonumber \\
\alpha &=&5/4,\ \ \ y\simeq 2800\   \nonumber \\
\alpha &=&1,\text{ \ \ \ \ \ no boundary}.  \nonumber
\end{eqnarray}
Since $\alpha =2$ is a vacuum solution, one could attempt construction of a
matching vacuum for the lower values of $\alpha $. However, the lack of a
boundary at $\alpha =1$ implies that the boundaries mark a change from
pressure to tension rather than a boundary to vacuum. This supports a
picture of an atmosphere which extends farther as the size of the observed
fluctuation decreases, all asymptotically approaching vacuum. We again
remark that asymptotic flatness applies only to the observer's $\alpha =2$
spacetime.

\subsection{THE SIERPINSKI GASKET}

These ideas can be applied to diffusive processes whose continuum solutions
are expected to envelope fractal behavior at smaller scales. As an example,
we will use some existing discussions of enveloped fractal structures in
flat spaces to discuss the approximate sizes of $\alpha $ and $\beta $ where 
$D(r)=D_{_0}r^{\beta -\alpha }=D_{_0}r^\Theta $. The numerical parameters of
interest are $\alpha ,$ the diffusivity factor $\Theta ,$ the Euclidean
embedding dimension $d_E$ in which the fractal substrate is embedded, and a
fractal dimension $d_F$ (which may not be $d_H$). Alexander and Orbach \cite
{a-o} and O'Shaughnessy and Procaccia \cite{osh-pro} provide some numerical
values of $d_F$ and $\Theta $ for various fractal processes. The fractal
dimension $d_F$ is defined through the diffusion equation which is written
as 
\[
\frac{\partial \Gamma }{\partial u}=\frac 1{r^{d_F-1}}\frac{\partial
(D_{_0}r^{-\Theta }r^{d_F-1}\partial _r\Gamma )}{\partial r} 
\]
In $d_E=3$, the fractal dimension for the Sierpinski gasket is $d_F=2$ and $%
\Theta =0.59.$ The corresponding values of $\alpha $ and $\beta $ are $%
\alpha =1,\ \beta =1.585.$ For $d_E=4$, the fractal dimension is $d_F=2.31$
and $\Theta =0.81$. We find $\alpha =1.32$ and $\beta =2.13$.

For all of the above examples, we could find a $u$-independent solution of
the diffusion equations as 
\begin{eqnarray*}
\Gamma &=&\Gamma _{0}r^{1-\beta }, \\
M &=&M_{0}+M_{1}r^{\alpha +1}.
\end{eqnarray*}
The mass scaling is set entirely by the metric parameter $\alpha $, while
the density depends on the scaling of the diffusivity.

\section{CONCLUSION}

Introducing a parameter, $\alpha $, into the extended Schwarzschild metric,
produces an atmosphere with processes that can be linked to a fractal mass
distribution. Specific values of $\alpha $ can be associated with particular
fractal substrates in a small scale region of the atmosphere. The continuous
solutions of an anomalous diffusion equation in the atmosphere envelope
smaller scale fractal processes. Anomalous diffusion can sample fractal
processes on a topological level, giving information about the connectivity
as well as the fractal dimension. Fluid parameters from the field equations
can be studied for a range of $\alpha ,$ allowing the effects of
fluctuations from one scale to another to be examined.

Other models are possible. For example, a fractal Schwarzschild cell could
be used in a universe model like the Lindquist-Wheeler model \cite{l-w},
perhaps describing the fractal galactic density distribution. If the
galactic distribution can be modeled by a single fractal process in which
both mass and density diffuse, then on time scales that are short compared
with the diffusion time, the mass would scale with the fractal dimension
while the scaling of the density would involve topological elements of the
fractal substrate. In this paper only a single level of scale reduction has
been studied. It is possible that an ensemble of solutions, a multifractal
set summed over $\alpha $, could be used to model more complex physical
systems.

Another direction for future work is a possible link between fractal
atmospheric structure and chaos. A variety of perturbations to a black hole
solution have been shown to lead to chaotic behavior, most often chaotic
particle orbits {\cite{rugh}}. For example, Bombelli and Calzetta \cite{b-c}
have shown that gravitational perturbations of the Schwarzschild solution
can produce chaotic orbits. In general, the effects of chaos range from
unusual production of gravitational waves \cite{c-f} to a possible
brightening of the black hole \cite{levin}. Orbiting particles with spin may
also exhibit chaotic behavior {\cite{s-m1},\cite{s-m2}}. A relation between
a phase space fractal dimension and chaos has been established for multi
black hole spacetimes \cite{d-f-c} and for black hole chaotic scattering
processes \cite{f-lar},\cite{dem-let}. We have considered an atmosphere that
has fractal structure in the physical spacetime. At one scale, the solution
can describe an ideal black hole, or with time dependence and space
dependence, a black hole with a two fluid atmosphere. The effect of
perturbations of this atmosphere and the effects of the physical fractal
structure on particle orbits is an interesting area for investigation.

\end{document}